\documentstyle[12pt]{article}
\newcommand{\AAA}{{\cal A}}

\newcommand{\vo}{\vec \omega}
\newcommand{\vn}{\vec \nabla}
\newcommand{\vr}{\vec r}

\newcommand{\be}{\begin{equation}}
\newcommand{\ee}{\end{equation}}
\newcommand{\ben}{\begin{eqnarray}\displaystyle}
\newcommand{\een}{\end{eqnarray}}
\newcommand{\refb}[1]{(\ref{#1})}

\begin{document}

{}~ \hfill\vbox{\hbox{hep-th/9707042}\hbox{MRI-PHY/P970716}}\break

\vskip 3.5cm

\centerline{\large \bf Dynamics of Multiple Kaluza-Klein 
Monopoles}

\centerline{\large \bf in $M$- and String Theory}

\vspace*{6.0ex}

\centerline{\large \rm Ashoke Sen\footnote{On leave of absence from 
Tata Institute of Fundamental Research, Homi Bhabha Road, 
Bombay 400005, INDIA}
\footnote{E-mail: sen@mri.ernet.in, sen@theory.tifr.res.in}}

\vspace*{1.5ex}

\centerline{\large \it Mehta Research Institute of}
 \centerline{\large \it   Mathematics and Mathematical Physics}

\centerline{\large \it Chhatnag Road, Jhusi, Allahabad 221506, INDIA}

\vspace*{4.5ex}

\centerline {\bf Abstract}

We analyse the world-volume theory of multiple Kaluza-Klein
monopoles in string and $M$- theory by identifying the appropriate
zero modes of various fields. The results are consistent with
supersymmetry, and all conjectured duality symmetries. In particular
for $M$-theory and type IIA string theory, the low energy dynamics
of $N$ Kaluza-Klein monopoles
is described by supersymmetric $U(N)$ gauge theory, and for
type IIB string theory, the low energy dynamics 
is described by a (2,0) supersymmetric field theory in (5+1)
dimensions with $N$ tensor multiplets and tensionless 
self-dual strings.  It is also argued
that for the Kaluza-Klein monopoles in heterotic string theory, the
apparently flat moduli space gets
converted to the moduli space of BPS monopoles in $SU(2)$
gauge theory
when higher derivative corrections to the supergravity equations 
of motion are taken into account. 

\vfill \eject

\baselineskip=18pt

\section{Introduction}

Among the brane solutions in $M$-theory and various string theories
are Kaluza-Klein monopoles\cite{SOR,GP}, 
which represent six branes in
$M$-theory compactified on a circle, and five-branes in various string
theories compactified on a circle. Various conjectured string
dualities relate these solitons to various other known solitons
in string theory. For example, in type IIA (IIB) string theory,
T-duality relates these to appropriate solitonic 
five branes in type IIB (IIA) string theory, whereas in
type IIA string theory and $M$-theory, U-duality relates the
Kaluza-Klein monopoles to appropriate D-branes in type IIA string 
theory. In this paper we shall explicitly test these duality
conjectures by comparing the low energy dynamics of multiple
Kaluza-Klein monopoles to that of various other string solitons. 
Dynamics of Kaluza-Klein monopoles have been analysed
previously in different contexts in 
refs.\cite{RUBACK,HULLNEW,KK,BERG,IMA}.

Kaluza-Klein monopole solutions\cite{SOR,GP} have the special property
that at a generic point in the moduli space, they represent 
completely non-singular solutions of the low energy supergravity
equations of motion. The curvature can be kept small everywhere
by taking the radius of the compact direction 
and the distance between different monopoles sufficiently
large, whereas the string coupling can be kept small everywhere
by keeping the asymptotic coupling constant small. This means
that at least at a generic point in the moduli space,
we can reliably analyse the low energy dynamics of these
monopoles using the weak coupling, low energy
supergravity equations of
motion. At special points in the moduli space, when two or more
monopoles coincide, stringy effects become important. As we
shall see, in this case the dynamics can be approximated by
string propagation in an appropriate ALE space, $-$ a problem
that has already been studied in other contexts. We shall find
that
the dynamics of the Kaluza-Klein monopoles, both at generic and
at special points in the moduli space, are consistent with
various duality conjectures.

Besides providing new tests of various duality conjectures,
the world volume theory of Kaluza-Klein monopoles can also be
used to study novel supersymmetric theories. For example, for
type IIB string theory, the world volume theory of $N$ Kaluza-Klein
monopoles is a (5+1)
dimensional theory with (2,0) supersymmetry and $N$ tensor multiplets,
and also contains self-dual strings which become tensionless
when the locations of some of the monopoles
coincide. This is closely related to the theories that have 
been used to define M(atrix)-theory\cite{MATRIX} on $T^4$ and
$T^5$\cite{ROS,SEIROS}. Although the solitonic five-branes
in type IIA
string theory have the same world-volume theory, describing the
theory in terms of Kaluza-Klein monopoles has the special
advantage that at a generic point in the moduli space the
string coupling can be kept small everywhere, and hence this
world volume theory can be described in terms of a weakly coupled
string theory.

\section{Review of Multiple Kaluza-Klein Monopole Solution}

The multiple Kaluza-Klein monopole solution is described by the
metric:
\be \label{e1}
ds^2 = -dt^2 + \sum_{m=5}^D dy^m dy^m + ds_{TN}^2\, ,
\ee
where $D$ is 9 for various string theories and 10 for $M$-theory,
$y^m$ denote the space-like world-volume coordinates on the
($D-4)$-brane represented by this solution,  and $ds_{TN}$ is the
metric of the Euclidean multi-centered
Taub-NUT space\cite{GIBHAW}:
\be \label{e2}
ds_{TN}^2 = U^{-1} (dx^4 + \vec \omega\cdot d\vr)^2 + U d\vr^2\, .
\ee
Here $x^4$ denotes the compact direction, and
$\vr \equiv(x^1, x^2, x^3)$ 
denotes the three spatial coordinates transverse to the brane.
$U$ and $\vo$ are defined as follows:
\be \label{e3a}
U = 1 + \sum_{I=1}^N U_I, \qquad \vo = \sum_{I=1}^N \vo_I\, ,
\ee
where,
\be \label{e4}
U_I={4m \over |\vr - \vr_I|}\, ,
\ee
and,
\be \label{e5}
\vn\times \vo_I = \vn U_I\, .
\ee
$m$ and $\vec r_I$ are parameters labelling the solution.
$\vec r_I$ can be interpreted as the locations of the
Kaluza-Klein monopoles in the transverse space. 
In order that the solutions are
free from conical singularities at $\vec r=\vec r_I$, $x^4$
must have periodicity $16\pi m$.

The multi-centered Taub-NUT space described by the
metric \refb{e2} supports $N$ linearly independent
normalizable self-dual harmonic two forms $\Omega_I$, given as
follows\cite{RUBACK}:
\be \label{e6}
\Omega_I = d \xi_I\, ,
\ee
where,
\be \label{e7}
\xi_I = U^{-1} U_I (dx^4 +\vo\cdot d\vr) - \vo_I\cdot d\vr\, .
\ee
Furthermore, as shown in \cite{RUBACK}, the two-forms defined
above satisfy a simple normalization condition:
\be \label{e7a}
\int \Omega_I\wedge \Omega_J = (16\pi m)^2 \delta_{IJ}\, .
\ee

\section{The World-volume Theory at a Generic Point in the Moduli
Space}

Let us now suppose that the original string / $M$- theory has a $p$
form field $A^{(p)}$. Generalizing the construction of
refs.\cite{KK,IMA} for the case of single monopole, we can define
$N$ different
$(p-2)$-form fields $\AAA^{(p-2)}_I$
on the world-volume through the decomposition:
\be \label{e12}
A^{(p)}(\vr, x^4, \vec y, t) = \sum_{I=1}^N \AAA_I^{(p-2)}(\vec y,
t)\wedge \Omega_I(\vr,x^4)\, .
\ee
Besides the fields obtained this way, there are also $3N$
world-volume scalar fields $\Phi^i_I$ ($1\le i\le 3$) associated
with the parameters $\vr_I$. These can be associated with
deformations of the metric following the procedure outlined in
ref.\cite{RUBACK}. This gives the following world-volume field
theories of Kaluza-Klein monopoles in different theories at a generic
point in the moduli space:

\subsection{$M$-theory} 

This theory contains a three form gauge
field $C_{MNP}$ in (10+1) dimensions. This gives $N$ vector
fields on the (6+1) dimensional world volume of the Kaluza-Klein
monopole. Together with the $3N$ scalar fields coming from the
translational zero modes, these fields describe the bosonic
fields of an $N=1$ supersymmetric $(U(1))^N$ gauge theory in (6+1)
dimensions. The metric on the moduli space spanned by the scalar
fields is determined by the inner product
$\int\Omega_I\wedge \Omega_J$\cite{RUBACK}.
{}From the moduli independence of the inner product given in
eq.\refb{e7a} we see that this metric is flat, as is 
expected for this particular supersymmetric field
theory\cite{GIBRECENT}.\footnote{The fermionic fields on the 
world-volume come
from the various fermionic zero modes, but we shall not carry out
the explicit analysis here, since the world-volume supersymmetry
uniquely fixes the fermionic field content and their
interactions.}

According to the conjectured duality between $M$-theory on $S^1$
and type IIA string theory, the Kaluza-Klein monopole in $M$-theory
is identified to the Dirichlet six-brane of type IIA string
theory. The world-volume dynamics of $N$ well separated Dirichlet
six branes in type IIA string theory is described by an N=1
supersymmetric $(U(1))^N$ gauge theory in (6+1) dimensions. Thus we
see that the dynamics of the Kaluza-Klein monopoles is in
agreement with the conjectured duality between $M$-theory and type
IIA string theory. 

\subsection{Type IIA string theory}

This theory contains a
three form gauge field $C_{MNP}$ and a one form field $A_M$ from
the Ramond-Ramond (RR) sector, and a two form field $B_{MN}$ from
the Neveu-Schwarz$-$Neveu-Schwarz (NS) sector. $C_{MNP}$ gives
$N$ vector fields 
whereas $B_{MN}$ gives $N$ scalar fields on the 
(5+1) dimensional world-volume of the monopole. Combining these
 with the
$3N$ scalar fields coming from the translational zero modes, we
get the bosonic field content of (1,1) supersymmetric $(U(1))^N$
gauge theory in (5+1) dimension. Actually, $N$ of these scalar
fields, associated with the modes of $B_{MN}$, are periodic
coordinates\cite{KK}.  
Thus the world volume field theory is more naturally
interpreted as a $(6+1)$ dimensional N=1 supersymmetric $(U(1))^N$
gauge theory, compactified on a circle of appropriate radius.
The $N$ scalar fields originating from $B_{MN}$ are 
then interpreted
as the components of the $(U(1))^N$ gauge fields along the compact
direction. 

In order to see how this result is compatible with the
predictions of duality, let us note that first of all, by a
T-duality transformation, these Kaluza-Klein monopoles are
related to solitonic five branes of type IIB theory compactified
on a circle $S^1$, with the five-branes being transverse to $S^1$.
The S-duality transformation of
type IIB string theory converts these solitonic five branes
to Dirichlet five branes transverse to $S^1$. Finally, by
a T-duality transformation, this can be mapped to a system of
Dirichlet six
branes in type IIA string theory on $S^1$
with the six branes wrapped on the compact direction. For
$N$ well separated six branes, the world-volume theory of this
system is given by an N=1 supersymmetric $(U(1))^N$ gauge theory in
(6+1) dimensions,
compactified on a circle. Thus we
see that the world volume theory of $N$ well separated
Kaluza-Klein monopoles in type IIA string theory is consistent
with the predictions of duality.

\subsection{Type IIB string theory}

Type IIB string theory
contains a four form field $D_{MNPQ}$ with self-dual field
strength, and a pair of two form gauge fields $B_{MN}$ and
$B'_{MN}$, one from the NS sector and one from the RR sector.
$D_{MNPQ}$ gives rise to $N$ self-dual two form gauge fields on the
(5+1) dimensional world volume, whereas $B_{MN}$ and $B'_{MN}$
give rise to $2N$ scalar fields on the world-volume. Combining
these with the $3N$ scalar fields from the translational zero
modes, we get the bosonic field content of $N$ tensor multiplets
of the (2,0) supersymmetric field theory in (5+1) dimensions. $2N$ of
the $5N$ bosonic fields, associated with the modes of $B_{MN}$
and $B'_{MN}$ represent periodic coordinates. 

In order to see how this is consistent with various duality
conjectures, we can make a T-duality transformation to
relate it to solitonic five branes of type IIA on $S^1$, with the
five branes being transverse to $S^1$. By using the duality
between $M$-theory on $S^1$ and type IIA string theory, we can
further relate it to a system of five branes in $M$-theory on
$T^2$, with the five branes being transverse to both coordinates
on $T^2$. The world volume theory of this system for $N$ well
separated five branes does indeed contain $N$ tensor multiplets
of the (2,0) supersymmetry algebra. The $5N$ bosonic fields
represent the five transverse coordinates of each of the $N$ five
branes. Since in this case two of the transverse coordinates are
compactified on circles, $2N$ of the $5N$ bosonic fields must 
represent periodic coordinates. Thus we see that the world volume
theory of multiple Kaluza-Klein monopoles in type IIB string
theory is also consistent with the predictions of duality.

\subsection{Heterotic String Theory}

This theory contains a
rank two anti-symmetric tensor field $B_{MN}$, which gives rise
to $N$ periodic bosonic fields. These fields, together with the
$3N$ bosonic fields associated with the translational zero modes,
constitute the field content of $N$ hypermultiplets of the (1,0) 
supersymmetry algebra in (5+1) dimensions. However,
unlike in the previous
cases, in this case the field content, together with the
requirement of supersymmetry, does not determine the low energy
effective action completely. One needs to specify the moduli
space of these hypermultiplets in order to 
completely describe this theory.
According to the result of low energy supergravity
theory, this moduli space is flat\cite{RUBACK}, and is given by
$(R^3\times S^1)^N/S_N$, where in the $I$th term in this product,
$R^3$ is labelled by the transverse
coordinates of the $I$th monopole, 
and $S^1$ is labelled by the component of the
$B_{MN}$ field along $\Omega_I$. $S_N$ denotes the effect of
symmetrization under permutations of different monopoles.
This result however is valid only for large radius $R$
of the compact direction
and large separation between the monopoles, since in this case the
curvature associated with the solution \refb{e1}, \refb{e2} is
small, and hence higher derivative stringy corrections to the low
energy supergravity equations of motion are negligible.
However, as we shall now argue, string world-sheet corrections
convert this into the moduli space of
$N$ BPS monopoles in $SU(2)$ gauge theory\cite{MAN,AH,GIBMAN}.

To see this, let us recall that a Kaluza-Klein monopole in the
heterotic string theory carries 1 unit of Kaluza-Klein monopole
charge, and $-1$ unit of $H$-monopole charge\cite{KK}.
(The $H$-monopole charge arises from the 
Lorentz Chern-Simons term in the expression for the field strength
of the anti-symmetric tensor field.)  
Now, starting from a situation where the radius of the compact 
direction is large, and hence the solutions \refb{e1}, \refb{e2}
are valid, consider slowly reducing
the radius (keeping the string coupling small) so that it comes
close to the self-dual radius. In this case, the low energy
effective theory describing heterotic string theory on $S^1$
contains a spontaneously broken $SU(2)$ gauge group\cite{NARAIN}. 
This theory
will contain BPS monopoles associated with this broken $SU(2)$. It
is easy to see that this BPS monopole carries precisely the same
quantum numbers as the Kaluza-Klein monopole. Hence we would
expect that as we slowly change the radius from large value to
the self-dual point, the moduli space of the Kaluza-Klein monopoles 
will evolve to that of BPS monopoles. (Note that since we can
keep the string coupling small throughout this process, the
moduli space approximation never breaks down.)

The question we would like to ask is: what
is the metric that interpolates between
the moduli space metric of the Kaluza-Klein monopoles and that of
BPS monopoles? We shall now show that it is consistent to assume
that the correct metric is that of the moduli space of BPS
monopoles, and the flat metric on the Kaluza-Klein monopole
moduli space emerges from this in appropriate limit when we
ignore higher derivative corrections to the supergravity
equations of motion.  Implicitly, this may be understood as 
follows.  The
moduli space of multiple BPS monopoles involves a scale factor,
which, in turn, is determined by the
inverse vacuum expectation value (vev)
of the Higgs field responsible
for the breaking of $SU(2)$ to $U(1)$.  If we take the limit where
this scale factor goes to zero ({\it i.e.} the vev of the
Higgs field goes to
infinity), keeping the separation between the monopoles fixed,
the metric on the moduli space of $N$ BPS monopoles 
approaches flat metric.
Now, since the Higgs vacuum expectation value
in the $SU(2)$ gauge theory is proportional to $(R-R^{-1})$ (in
units in which $R=1$ denotes the self-dual radius), we see that
for large $R$ the Higgs vacuum expectation becomes large, and the
scale factor indeed goes to zero. Thus in this limit we expect
the moduli space metric of BPS monopoles to approach flat metric,
as we have found for the Kaluza-Klein monopoles.

To see this more explicitly,
let us analyse the components of the metric in the cartesian coordinate
system in which the metric for well separated monopoles approaches the
identity matrix. In this coordinate system, the various
components of the metric on the moduli space of Kaluza-Klein
monopoles are naturally functions of $(\vr_I-\vr_J)/R\equiv
\vec\xi_{IJ}$ if we restrict ourselves to using the low energy
supergravity equations. The Planck scale or the string scale
never enters the calculation since Einstein's equations without
stringy corrections and without source terms are scale
invariant. The result of \cite{RUBACK} shows that the 
metric is identity matrix, {\it i.e.} the
metric components do not depend on $\vec\xi_{IJ}$. On the other hand,
components of the metric on the moduli space of BPS monopoles 
are functions of the combination
$(R-R^{-1})(\vr_I-\vr_J)= (R^2-1) \vec\xi_{IJ}$ for large
$(R-R^{-1})|\vr_I-\vr_J|$, and these components differ 
from those of the
identity matrix by terms of order
$1/((R^2-1)\xi_{IJ})$ when $(R^2-1)\xi_{IJ}$ are 
large\cite{GIBMANT}. Straightforward
dimensional analysis shows us that these terms must be
accompanied by the square of the string length in the numerator, and
hence
are expected to show up in the computation of the metric of 
Kaluza-Klein monopoles only when higher derivative corrections
to the supergravity equations of motion are taken into account.
In the limit when the string length goes to zero, the metric
reduces to the flat metric, as is the case for the moduli space
of Kaluza-Klein monopoles in the supergravity theory.

Thus we see that flat metric on the moduli space of Kaluza-Klein
monopoles as derived from low energy supergravity theory
is perfectly consistent with the idea that the correct
moduli space is that of multiple BPS monopoles in $SU(2)$ gauge
theory, scaled by $(R-R^{-1})$. Since the later
moduli space has all the properties required for being consistent
with supersymmetry and various other symmetries, and correctly
interpolates between the known moduli spaces near the self-dual
radius and for large radius of the compact direction, we propose
that this is the corrrect moduli space.

\section{Singular Points in the Moduli Space}

Let us now turn to special points in the moduli space where the
locations of some of the monopoles coincide. We shall consider
the extreme case where the locations of all the $N$ monopoles
coincide; any other configuration can be analyzed in an identical
manner. This corresponds to taking all the $\vr_I$'s appearing in
eq.\refb{e4} to be close to each other. For definiteness, we
shall take all of them to be close to $\vr=0$. In this case, the
metric near $\vr=0$ can be approximated by:
\be \label{e2a}
ds_{TN}^2 \simeq U^{\prime -1} (dx^4 + \vec \omega'\cdot d\vr)^2 + 
U' d\vr^2\, .
\ee
where,
\be \label{e13}
U' = \sum_{I=1}^N {4m\over |\vr -\vr_N|}, \qquad
\vn\times\vo' = \vn U'\, .
\ee
But this is precisely the metric on the ALE 
space\cite{GIBHAW,EGU}. When
all the $\vr_N$'s approach the point $\vr=0$, this space develops
an $A_{N-1}$ type singularity, and we get new massless particles
/ tensionless strings
in different string theories / $M$- theory from 
branes wrapped around
collapsed two cycles associated with the singularity. We shall
now analyze the dynamics of massless particles / tensionless
strings in different theories near this singular point and show
that in each case the result is in agreement with the predictions
of duality. (This analysis is somewhat reminiscent of
that in ref.\cite{BERSAD}).

\subsection{$M$-theory}

$M$-theory on an ALE space with
$A_{N-1}$ singularity is expected to have enhanced $SU(N)$ gauge
symmetry\cite{WITTD}. Since for well separated monopoles the
world volume theory was an N=1 supersymmetric $(U(1))^N$ gauge
theory in (6+1) dimensions, for coincident monopoles it must be
described by an N=1 supersymmetric $U(N)$ gauge theory. We had
seen earlier that this system is related by duality to a system
of $N$ six branes in type IIA string theory. For coincident six
branes, the world volume theory of this system is indeed
described by an N=1 supersymmetric $U(N)$ gauge theory. Thus we
see that the enhanced gauge symmetry on the world volume of
coincident Kaluza-Klein monopoles in $M$-theory is also consistent
with the conjectured duality between $M$-theory on $S^1$ and type
IIA string theory.

\subsection{Type IIA String Theory}

Type IIA string theory on an ALE space with $A_{N-1}$ singularity
is also expected to have enhanced $SU(N)$ gauge
symmetry\cite{WITTD}. Combining this with the result for well
separated monopoles, we see that the world-volume theory of $N$
coincident Kaluza-Klein monopoles in type IIA string theory is
given by an $N=1$ supersymmetric $U(N)$ gauge theory in (6+1)
dimensions compactified on a circle. Our earlier analysis shows
that this system is dual to a system of $N$ Dirichlet
six branes of type IIA string theory wrapped on a circle. This
indeed develops $U(N)$ enhanced gauge symmetry for coincident six
branes, showing that the enhancement of gauge symmetry for
coincident Kaluza-Klein monopoles is consistent with the duality
conjecture.

\subsection{Type IIB String Theory}

Type IIB string theory on an ALE space with $A_{N-1}$ singularity
is expected to have no enhanced gauge symmetry, but tensionless
self-dual strings\cite{WITTTEN} whose charges under the 
$N$ tensor fields lie along the
root vectors of $U(N)$. These tensionless strings arise from
three branes wrapped along collapsed two cycles in the ALE space.
To see if this is consistent with duality, recall that the
Kaluza-Klein monopoles in type IIB string theory on $S^1$ are
related by duality to a system of five-branes in $M$-theory 
on $T^2$, with the five branes being transverse to both
directions of $T^2$. When the five branes coincide, we do get
tensionless strings from membranes stretched between
five-branes\cite{STROM,TOWN}. Hence again our result is
consistent with the conjectured duality between $M$-theory on
$S^1$ and type IIA string theory.

We also have strings of finite
tension in the world-volume theory of the five-branes in 
$M$-theory on $T^2$. These originate from membranes
stretched between the five branes which, instead of following
the direct path, wrap once or more around the compact
directions\cite{SEIROS}. One might ask: where do these finite
tension strings on the world-volume theory originate in the 
Kaluza-Klein monopoles in type IIB string theory? A
straightforward application of duality relations shows us that in
the world-volume of the Kaluza-Klein monopoles in
type IIB string theory, these new strings carry the same quantum
number as bound states of a three brane wrapped along a two
cycle, and
elementary strings and / or D-strings. Since the electric and
magnetic flux of the world-volume gauge field on the three brane
act as sources of $B_{MN}$ and $B'_{MN}$ 
respectively\cite{WITTSM,SCHMID}, we see
that in the world volume theory of Kaluza-Klein monopoles in
type IIB theory, the new finite tension strings can be 
identified with three
branes wrapped along collapsed two cycles, together with quantized
electric / magnetic flux of the three brane world-volume gauge
field through the two cycle.

\subsection{Heterotic String Theory}

Finally we turn to coincident Kaluza-Klein monopoles in the
heterotic string theory. Our earlier assertion that the moduli
space of these monopoles is identical to that of multiple BPS
monopoles in $SU(2)$ gauge theory implies that there is no
singularity in the moduli space even for coincident monopoles,
{\it i.e.} even when we put the heterotic string theory on a
singular ALE space. It will be interesting to find an independent
verification of this result.

\section{Conclusion}

In this paper we have studied the dynamics of multiple
Kaluza-Klein monopoles in $M$-theory, and various string
theories, for well separated as well as coincident monopoles.
In each case, we have found that the dynamics is consistent with
all known duality conjectures involving string / $M$- theory.
Dynamics of multiple Kaluza-Klein monopoles in type IIB string
theory gives a new realization of the (2,0) supersymmetric
theory in (5+1) dimensions with multiple tensor multiplets and
tensionless self-dual strings. This theory in appropriate
limits might be 
relevant for describing M(atrix)-theory on $T^4$ and $T^5$.
We have also argued that the 
dynamics of $N$ Kaluza-Klein monopoles in heterotic string
theory is described by a (1,0) supersymmetric field theory in
(5+1) dimensions, with moduli space given by the moduli space of
$N$ BPS monopoles in $SU(2)$ gauge theory.


\begin{thebibliography}{99}

\bibitem{SOR}
R. Sorkin, Phys. Rev. Lett. {\bf 51} (1983) 87.

\bibitem{GP}
D. Gross and M. Perry, Nucl. Phys. {\bf B226} (1983) 29.

\bibitem{RUBACK}
P. Ruback, Comm. Math. Phys. {\bf 107} (1986).

\bibitem{HULLNEW}
C. Hull, [hep-th/9705162].

\bibitem{KK}
A. Sen, [hep-th/9705212].

\bibitem{BERG}
E. Bergshoeff, B. Janssen and T. Ortin, [hep-th/9706117].

\bibitem{IMA}
Y. Imamura, [hep-th/9706144].

\bibitem{MATRIX}
T. Banks, W. Fischler, S. Shenker and L. Susskind,
Phys. Rev. {\bf D55} (1997) 112
[hep-th/9610043]. 

\bibitem{ROS}
M. Rozali, [hep-th/9702136].

\bibitem{SEIROS}
M. Berkooz, M. Rozali and N. Seiberg, [hep-th/9704089].

\bibitem{GIBHAW}
S. Hawking, Phys. Lett. {\bf 60A} (1977) 81; \\
G. Gibbons and S. Hawking, Comm. Math. Phys. {\bf 66} (1979) 291.

\bibitem{GIBRECENT}
G. Gibbons, G. Papadopoulos and K. Stelle, [hep-th/9706207], and 
references therein.


\bibitem{MAN}
N. Manton, Phys. Lett. {\bf 110B} (1982) 54; Phys. Lett. {\bf
154B} (1985) 397.

\bibitem{AH}
M. Atiyah and N. Hitchin, Phys. Lett. {\bf 107A} (1985) 21;
Phil. Trans. R. Soc. Lond. {\bf A315} (1985) 459;
The Geometry and Dynamics of Magnetic Monopoles, Princeton Univ.
Press (1988).

\bibitem{GIBMAN}
G. Gibbons and N. Manton, Nucl. Phys. {\bf B274} (1986) 183.

\bibitem{NARAIN}
K. Narain, Phys. Lett. {\bf 169B} (1986) 41; \\
K. Narain, H. Sarmadi and E. Witten, Nucl. Phys. {\bf B279}
(1987) 369.

\bibitem{GIBMANT}
G. Gibbons and N. Manton, Phys. Lett. {\bf B356} (1995) 32
[hep-th/9506052].

\bibitem{EGU}
T. Eguchi, B. Gilkey and A. Hanson, Phys. Rep. {\bf 66} (1980)
213, and references therein.

\bibitem{BERSAD}
M. Bershadsky, V. Sadov and C. Vafa, Nucl. Phys. {\bf B463} (1996)
398 [hep-th/9510225]; \\
H. Ooguri and C. Vafa, Nucl. Phys. {\bf B463} (1996) 55
[hep-th/9511164].

\bibitem{WITTD}
E. Witten, Nucl. Phys. {\bf B443} (1995) 85 [hep-th/9503124].

\bibitem{WITTTEN}
E. Witten, [hep-th/9507121].

\bibitem{STROM}
A. Strominger, Phys. Lett. {\bf B383} (1996) 44 [hep-th/9512059].

\bibitem{TOWN}
P. Townsend, Phys. Lett. {B373} (1996) 68 [hep-th/9512062].

\bibitem{WITTSM}
E. Witten, Nucl. Phys. {\bf B460} (1996) 335 [hep-th/9510135].

\bibitem{SCHMID}
R. Leigh, Mod. Phys. Lett. {\bf A4} (1989) 2767; \\
C. Schmidhuber, Nucl. Phys. {\bf B467} (1996) 146
[hep-th/9601003]; \\
M. Green and M. Gutperle, Phys. Lett. {\bf B377} (1996) 28
[hep-th/9602077].

\end{thebibliography}
\end{document}